\documentclass[10pt,twocolumn]{article}
\usepackage{times}
\usepackage{graphicx}
\usepackage{url}
\usepackage{mathfont}

\mathcode`O="724F

\makeatletter
\def\@begintheorem#1#2{\it \trivlist \item[\hskip \labelsep{\bf #1\ #2.\
}]}
\def\@opargbegintheorem#1#2#3{\it \trivlist
      \item[\hskip \labelsep{\bf #1\ #2\ (#3).}]}
\makeatother

\renewcommand{\topmargin}{-.50in}

\newtheorem{theorem}{Theorem}

\newtheorem{lemma}{Lemma}

\newcommand{\qed}{~$\vrule width.15cm height.2cm depth0cm$ \medbreak}
\newenvironment{proof}{\noindent{\bf Proof sketch: }}{\qed}

\DeclareSymbolFont{AMSb}{U}{msb}{m}{n}
\DeclareSymbolFontAlphabet{\Bbb}{AMSb}
\def\R{\ensuremath{\Bbb R}}
\def\E{\ensuremath{\Bbb E}}

\newcommand{\OCfigure}[3]{\begin{figure*}
 \center{\vbox to #3 {\vfil
	 \includegraphics[height=#3]{#1.eps}}}
 \caption{#2}
 \label{#1}
\end{figure*}}

\begin{document}

\title{M\"obius-Invariant Natural Neighbor Interpolation}
\author{Marshall Bern\thanks{PARC, 3333 Coyote Hill Rd., Palo 
Alto, CA 94304}\and
David Eppstein\thanks{UC Irvine, Dept. Inf. \& Comp. Sci., Irvine, CA 
92697. Work done in part while visiting PARC
and supported in part by NSF grant CCR-9912338.}}
\date{ }

\maketitle

\begin{abstract}
We propose an interpolation method that is invariant under
M\"obius transformations; that is, interpolation followed
by transformation gives the same result as transformation followed
by interpolation.   
The method uses natural (Delaunay) neighbors, but weights
neighbors according to angles formed by Delaunay circles. 
\end{abstract}

\section{Introduction}

An {\it inversion\/} of $\R^2 \cup \{ \infty \}$  
is a bijection generated by a circle, that maps circles to circles
and preserves angles between pairs of curves. 
The circle with center $c$ and radius $r$
gives an inversion as follows:  each point $p$ along a ray with origin $c$
is mapped to another point $p'$ along the same ray, so that the
product of the distances $|cp|$ and $|cp'|$ is $r^2$. 
Thus points on the circle remain fixed
and the center $c$ maps to $\infty$.
The set of products of inversions forms a group, the group of
{\it M\"obius transformations\/} on Euclidean space $\E^2$.

M\"obius transformations arise in data visualization. 
For example, hyperbolic browsing~\cite{Lamping,Munzner}
uses the Poincar\'e model of the hyperbolic plane to view
large graph structures.
A change of focus in a hyperbolic browser is given by an
isometry of hyperbolic space corresponding to a M\"obius transformation
of $\E^2$ that maps the unit disk to itself~\cite{Iversen}.
Brain mapping~\cite{Hurdal} 
flattens a convoluted surface by mapping it quasiconformally to a disk;
subsequent navigation in the brain map can be accomplished by hyperbolic browsing.
For these applications, 
it is reasonable to ask for computational primitives
specially designed for compatibility with M\"obius transformations.
We previously considered the problem of finding an optimal M\"obius
transformation for viewing hyperbolic or spherical data~\cite{Bern&01}.
In this paper we give a M\"obius-invariant method of extending
a function defined only at discrete sample points to a continuous
function defined everywhere on their convex hull. 

\section{Natural Neighbors}
Let $S = \{ s_1, s_2, \dots , s_n \}$ be a set of points in $\E^2$,
and let $s$ be another point in the convex hull of $S$.
Assume we have real- or complex-valued ``elevation'' $z_i$ for each $s_i$,
and we would like to define an elevation $z = f(s)$ at $s$, such
that $f$ is a continuous function of $s$ and  $f(s_i) = z_i$.

Natural neighbor interpolation~\cite{Sibson}
sets $f(s)$ to be a convex combination of the $z_i$ values at the
{\it natural neighbors\/} (also known as Delaunay or Vorono{\"\i } neighbors) of $s$.  
Let $V$ denote the Vorono{\"\i } cell  of $s$ in the 
Vorono{\"\i } diagram of $S \cup \{ s \})$
(that is, the set of points in $\E^2$ closer to $s$ than to any $s_i$),
and let $V_i$ denote 
the Vorono{\"\i } cell  of $s_i$ in the Vorono{\"\i } diagram of $S$.  
Then $f(s) = \sum_i w_i z_i$ where $w_i$ is the fraction of
$V$ covered by $V_i$. See Figure~\ref{vorfigs}(a).
Sibson~\cite{Sibson} showed that the weights $w_i$
act as a sort of local coordinate system for $s$ in the sense
that if $s = (x,y)$ and $s_i = (x_i, y_i)$, then
$x_i = \sum_i w_i x_i$
and 
$y_i = \sum_i w_i y_i$.
Sibson's theorem also implies that natural neighbor 
interpolation reconstructs linear functions, that is,
if $z_i = ax_i + by_i + c$ for each $i$ 
and some $a$, $b$, and $c$, then $z = ax + by + c$.

Natural neighbor interpolation is a logical starting point
for M\"obius-invariant interpolation, because  
the notion of Vorono{\"\i } neighbor can be made  
invariant under M\"obius transformations.  
Two sites $s_i$ and $s_j$ are {\it extended  Vorono{\"\i } neighbors\/}
if there is a circle $C$ through $s_i$ and $s_j$ 
bounding either an empty disk or empty disk complement.
After an inversion, $C$ or its
complement remains empty
and thus $s_i$ and $s_j$ are still neighbors.

\OCfigure{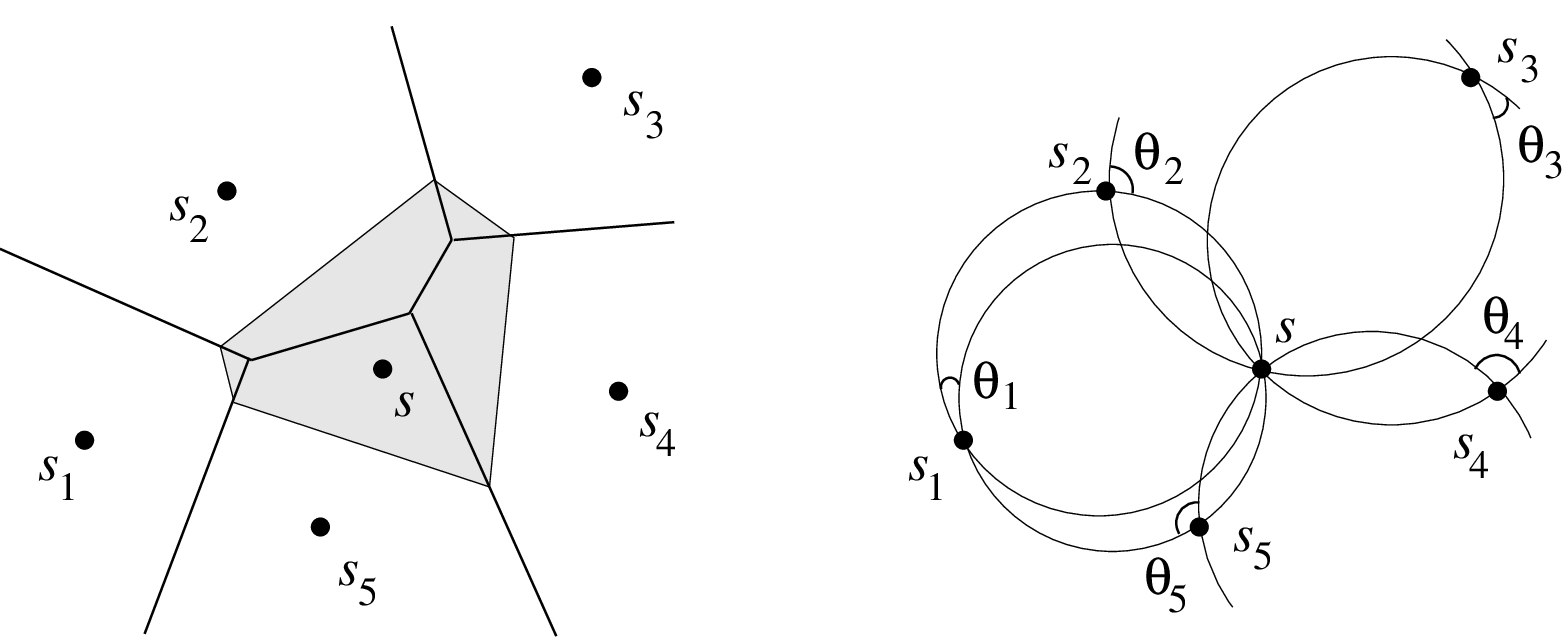}{(a)  Classical natural neighbor interpolation
weights neighbors according to fractions of the area of the Vorono{\"\i } cell of $s$.
(b) M\"obius-invariant natural neighbor interpolation weights
neighbors according to lune angles.}{1.8in}

\section{What Weights?}

What do we do with the weights $w_i$?
Area changes under M\"obius transformations, so we must find
an invariant alternative:  angle.
The angle most obviously associated with natural neighbor $s_i$ is the
angle at the {\it lune\/} formed by maximal empty circles,
as shown in Figure~\ref{vorfigs}(b). 
We use the exterior lune angles $\theta_i$, because the
sum of these angles is fixed: $\sum_i \theta_i = 2\pi$.
(One way to see this fact is to perform an 
inversion that takes $s$ to $\infty$; 
the exterior lune angles become the exterior angles around the
convex hull of the transformed point set.)
For invariance, we can simply set $w_i$ to be proportional
to any function of angle, $w(\theta_i)$. 
But is there some especially natural choice for $w()$? 

First, we would like the interpolation to be continuous as $s \to s_i$, so
we require $w(\theta_i) \to \infty$ as $\theta_i \to \pi$. 
Second, we would like the interpolation to reconstruct some class of functions, 
analogous to linear functions in Sibson's method. 
Harmonic functions (solutions to Laplace's 
equation $\nabla^2 f = 0$) would be high on anyone's wish list, because
of numerous applications (heat, electrostatics) and 
because the family of harmonic functions is
itself invariant under M\"obius transformations.
The values of a harmonic function on a discrete point set, however,
do not determine the function, so we cannot hope for a property as strong
as Sibson's theorem. 
We instead ask for a property that holds only in the limit.

\begin{lemma}
Let $s_1, s_2, \dots , s_n$ be a set of points on a circle $C$, and let $\epsilon$ 
be the maximum angle of arc between a pair of successive $s_i$'s. 
Let $s$ be a point interior to $C$, and let $\theta_i$ be
the exterior lune angle between $s$ and $s_i$ as above.
Now let $f$ be a harmonic function defined on the closed disk bounded by $C$.
Then as $\epsilon \to 0$,
$\sum_i \theta_i \, f(s_i) \,\to\,  f(s)$.
\end{lemma}

\begin{proof}
There is a M\"obius transformation 
that puts the image of $s$ at the center of the image of circle $C$. 
Now the harmonic measure of an arc of a disk with respect
to the disk center is proportional to its arc angle
(see for example~\cite{Pommerenke}).
It is not hard to confirm that in the limit of small arcs, the arc angles 
are the same as the exterior lune angles. 
Thus $\sum_i \, \theta_i f(s_i)$ is the sum over small arcs of 
harmonic measure times the value of $f()$ at a point in the arc; this
sum converges to the limit $\,f(s)$.
\end{proof}

Thus in summary we would like the weighting function 
$w(\theta_i)$ to go to infinity as $\theta_i \to \pi$
and $w(\theta_i)/\theta_i$ to go to a constant as $\theta_i \to 0$. 
The most obvious choice is to make $w(\theta_i)$ proportional to
$\tan(\theta_i/2)$.
We normalize the weights so that they sum to one, and thus the
actual weights are $w_i = (1/W) \tan(\theta_i/2)$, where
$W = \sum_i \tan(\theta_i/2)$.

\begin{theorem}
Natural neighbor interpolation using weights proportional
to $\tan(\theta_i/2)$ gives a continuous function $f(s)$
that interpolates the elevations at sample points
and, in the limit of dense samples on a circle,
reconstructs harmonic functions.
\end{theorem}

Open questions include:
Is there a nice $C^1$ continuous version? 
(One idea is to weight by $\tan^2(\theta_i/2)$, but
this loses the harmonic function property.) 
What about higher-dimensional M\"obius-invariant interpolation?

\bigskip
\noindent
{\bf Acknowledgments: }
We would like to thank Steve Vavasis for
helpful conversations.

\end{document}